\documentclass[aps,prd,superscriptaddress,showpacs,preprintnumbers,10pt]{revtex4}
\usepackage{graphicx,color,here}
\newcommand{\be}{\begin{equation}}
\newcommand{\ee}{\end{equation}}
\newcommand{\bea}{\begin{eqnarray}}
\newcommand{\eea}{\end{eqnarray}}

\newcommand{\bfP}{\mbox{\boldmath $P$}}

\def\lsim{\mathrel{\rlap{\lower4pt\hbox{\hskip1pt$\sim$}}\raise1pt\hbox{$<$}}}
\def\gsim{\mathrel{\rlap{\lower4pt\hbox{\hskip1pt$\sim$}}\raise1pt\hbox{$>$}}}
\def\nostrocostruttino#1\over#2{\mathrel{\mathop{\kern 0pt \rlap
{\hbox{$#1$}}} \hbox{\kern-.135em $#2$}}}

\def\xp{x^\prime}
\def\zp{z^\prime}

\def\Vec#1{{\bf #1}}

\def\D{{\mathrm d}}
\def\E{{\mathrm e}}
\begin{document}
\title{A systematic phenomenological study of the $\cos 2 \phi$ asymmetry 
in unpolarized semi--inclusive DIS}

\author{Vincenzo Barone}
\affiliation{Di.S.T.A., Universit\`a del Piemonte Orientale
``A. Avogadro'', \\ and INFN, Gruppo Collegato di Alessandria, 15100
Alessandria, Italy}
\author{Alexei Prokudin\footnote{Email address: \texttt{prokudin@to.infn.it}}}
\affiliation{Di.S.T.A., Universit\`a del Piemonte Orientale
``A. Avogadro'', \\ and INFN, Gruppo Collegato di Alessandria, 15100
Alessandria, Italy}
\affiliation{Dipartimento di Fisica Teorica, Universit\`a di Torino, 
             Via P. Giuria 1, I-10125 Torino, Italy}

\author{Bo-Qiang Ma\footnote{Email address: \texttt{mabq@phy.pku.edu.cn}}}
\affiliation{School of Physics and State Key Laboratory of Nuclear
Physics and Technology,\\ Peking University, Beijing 100871, China}


\begin{abstract}
We study the $\cos 2 \phi$ azimuthal
asymmetry in unpolarized semi-inclusive DIS, taking 
into account both the perturbative contribution (gluon emission and splitting) 
and the non perturbative effects arising from intrinsic transverse motion and 
transverse spin of quarks. In particular we explore 
the possibility to extract from $\langle \cos 2 \phi \rangle$ 
some information about the 
Boer--Mulders function $h_1^{\perp}$, which represents 
a transverse--polarization asymmetry of quarks inside 
an unpolarized hadron. 
Predictions are presented for the HERMES, COMPASS and JLab kinematics, 
where $\langle \cos 2 \phi \rangle$ is dominated 
by the 
kinematical higher--twist contribution, and turns to be 
of order of few percent. 
We show that a larger asymmetry in $\pi^-$ production, compared 
to $\pi^+$ production, would represent a signature
of the Boer--Mulders effect.  

\end{abstract}


\maketitle

\section{Introduction}

Transverse spin and transverse momentum of quarks are  
by now universally recognized as two essential ingredients 
of the structure of hadrons. Their correlations are described 
by a number of $k_T$--dependent distribution 
functions which
give rise to various observables in hard hadronic 
processes \cite{Barone:2001sp}. 
Among  these distributions a special role is played by the 
Sivers function $f_{1T}^{\perp}(x, k_T)$ 
\cite{Sivers:1989cc,Sivers:1990fh}, 
which represents an azimuthal asymmetry of unpolarized quarks 
inside a transversely polarized hadron, and by
its chirally--odd partner $h_1^{\perp}(x, k_T)$, 
the Boer--Mulders function \cite{Boer:1997nt}, 
which represents a transverse--polarization asymmetry of quarks 
inside an unpolarized hadron. 
While the Sivers function is responsible of 
single--spin asymmetries in transversely polarized 
semi--inclusive DIS (SIDIS), 
the Boer--Mulders function  
 generates azimuthal asymmetries in {\em unpolarized} 
processes. 
Boer \cite{Boer:1999mm} suggested in fact that 
$h_1^{\perp}$ could explain the large  
$\cos 2\phi$ asymmetries observed in unpolarized $\pi N$ 
Drell-Yan processes \cite{Falciano:1986wk,Guanziroli:1987rp}. 
This conclusion was confirmed by more refined 
model calculations in \cite{Lu:2004hu,Lu:2005rq}. An even 
larger effect is expected in $p \bar p$ Drell-Yan 
production \cite{Barone:2007}, a process to be studied by the  
PAX experiment at GSI-HESR \cite{Barone:2005pu}.

A $\cos 2 \phi$ asymmetry also occurs 
in unpolarized SIDIS, where it was 
already investigated at high $Q^2$ by the 
EMC experiment \cite{Arneodo:1987} and by ZEUS \cite{Breitweg:2000qh}, 
and is now being measured 
in the low--medium $Q^2$ region by many 
experimental collaborations (HERMES, COMPASS, JLAb). 
In SIDIS the Boer--Mulders distribution 
couples to a fragmentation function, 
the Collins function $H_1^{\perp}$ \cite{Collins:1992kk}, 
which describes the fragmentation 
of transversely polarized quarks into polarized
hadrons. 
The Boer--Mulders mechanism, however, is not the only source of 
a $\cos 2 \phi$ asymmetry in SIDIS. Two other contributions 
to this asymmetry arise from non--collinear kinematics at order $k_T^2/Q^2$
(the so--called Cahn effect)  
\cite{Cahn:1978se,Cahn:1989yf}, and   
 perturbative gluon radiation (i.e. order $\alpha_s$ 
QCD processes) \cite{Georgi:1978,Mendez:1978,Koenig:1982,Chay:1992}.  
It is clear that in order to get a correct understanding
of the experimental results 
on the $\cos 2 \phi$ asymmetry, and possibly to determine the 
Boer--Mulders function, which is 
one of the ultimate goals of  
global fits of transverse spin data, 
all contributing effects should 
be taken into account and reliably estimated (previous works 
\cite{Oganessyan:1998,Gamberg:2003},  
including a preliminary study of ours \cite{Barone:2006}, considered 
only some of these effects). 
The purpose 
of this paper is indeed to investigate 
the three sources 
(Boer--Mulders, kinematical higher twist, order--$\alpha_s$ 
perturbative QCD) of the $\cos 2 \phi$ asymmetry in 
leptoproduction, and to evaluate their contributions 
in the HERMES, COMPASS and JLab regimes. 
In particular, we will identify 
some signatures of the Boer--Mulders effect
and the most advantageous experimental conditions
to study it (for an early attempt in this 
direction see \cite{Gamberg:2003hf}). 
More generally, the phenomenological frame 
we will establish here is intended to be an aid in  
interpreting the results of the future measurements 
of $\langle \cos 2 \phi \rangle$. To this aim, 
we tried to reduce the model dependence 
of our analysis to the minimum: thus $h_1^{\perp}$ is  
parametrized 
using a simple relation with the Sivers function 
based on the impact--parameter approach   
and on lattice results, and borrowing $f_{1T}^{\perp}$ 
from a recent fit to SIDIS data.

\section{Theoretical framework}
\label{theory}

\begin{figure}[t]
\includegraphics[width=0.75\textwidth]
{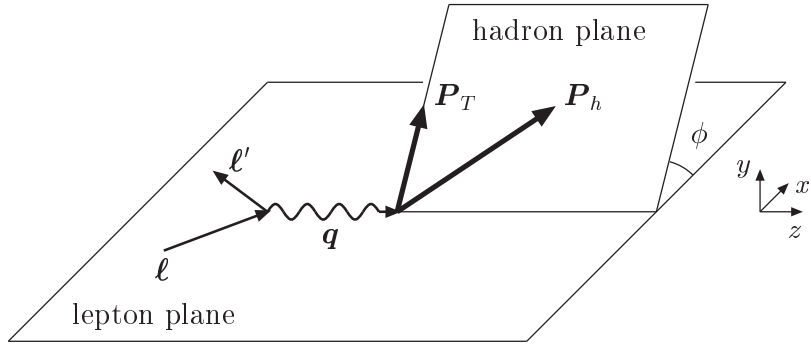}
\caption{\label{plane} Lepton and hadron planes in semi-inclusive 
deep inelastic scattering.
}
\end{figure}

The process we are interested in is unpolarized SIDIS:
\begin{equation}
l (\ell) \, + \, p (P) \, \rightarrow \, l' (\ell')
\, + \, h (P_h) \, + \, X (P_X)\,.
\label{sidis}
\end{equation}
The SIDIS cross section is expressed in terms of
the invariants
\begin{equation}
x = \frac{Q^2}{2 \, P \cdot q}, \;\;\;
y =  \frac{P \cdot q}{P \cdot \ell} ,
\;\;\;
z = \frac{P \cdot P_h}{P \cdot q}\,,
\end{equation}
where $ q = \ell - \ell'$ and $Q^2 \equiv - q^2$.
We choose the reference frame where
the virtual photon and the target proton are collinear
and directed along the $z$ axis, with the 
photon moving in the positive $z$ direction 
(Fig.~\ref{plane}). We denote by $\Vec k_T$ the transverse
momentum of the quark inside the proton, and by $\Vec P_T$ the
transverse momentum of the hadron $h$. The transverse momentum
of $h$ with respect to the direction of the fragmenting
quark will be called $\Vec p_T$. All azimuthal angles
are referred to the lepton scattering plane
(we call $\phi$ the azimuthal angle of the hadron $h$, 
see Fig.~\ref{plane}).

Taking the intrinsic motion of quarks into account, the
$\phi$--symmetric part of the 
SIDIS differential cross section reads at zero-th 
order in $\alpha_s$
\begin{eqnarray}
& &  \frac{\D^5 \sigma^{(0)}_{\rm sym}}{\D x \, \D y \, \D z \, \D^2 \Vec P_T}
 =  \frac{2 \pi \alpha_{\rm em}^2 s}{Q^4} \, \sum_a
e_a^2 \, x [1 + (1 - y)^2] \nonumber \\
& &  \hspace{1cm} \times \, \int \D^2 \Vec k_T \,
\int \D^2 \Vec p_T \,
\delta^2 (\Vec P_T - z \Vec k_T - \Vec p_T)
\, f_1^a (x, k_T^2) \,
D_1^a (z, p_T^2) \,,
\label{cross1}
\end{eqnarray}
where $f_1^a (x, k_T^2)$ is the unintegrated
number density of quarks of flavour $a$ and
$D_1^a (z, p_T^2)$ is the transverse-momentum
dependent fragmentation function of quark $a$ into
the final hadron. The non-collinear factorization
theorem for SIDIS has been proven
by Ji, Ma and Yuan \cite{Ji:2004xq} for
$P_{T} \ll Q$.

Long time ago Cahn \cite{Cahn:1978se,Cahn:1989yf} pointed out 
that the non-collinear 
kinematics generates a $\cos 2 \phi$ contribution
to the unpolarized SIDIS cross section, which
has the form
\begin{eqnarray}
& & \left. \frac{\D^5 \sigma^{(0)}_{\rm C}}{\D x \, 
\D y \, \D z \, \D^2 \Vec P_T}
\right \vert_{\cos 2 \phi}
 =  \frac{8 \pi \alpha_{\rm em}^2 s}{Q^4} \, \sum_a
e_a^2 \, x (1 - y) \nonumber \\
& & \hspace{2cm} \times \,
 \int \D^2 \Vec k_T \, \int \D^2 \Vec p_T
\,\delta^2 (\Vec P_T - z \Vec k_T - \Vec p_T)
\nonumber \\
& & \hspace{2cm} \times \,
\frac{2 \, (\Vec k_T \cdot \Vec h)^2 -  k_T^2}{Q^2} \,
f_1^a (x, k_T^2) \,
D_1^a (z, p_T^2)\, \cos 2 \phi \,,
\label{cross2}
\end{eqnarray}
 where $\Vec h \equiv \Vec P_T/P_T$. Notice that this contribution
is of order $k_T^2/Q^2$, hence it is a kinematical higher twist effect.

The second $k_T$-dependent source of the $\cos 2 \phi$
asymmetry involves the Boer-Mulders distribution $h_1^{\perp}$
coupled to the Collins fragmentation function $H_1^{\perp}$
of the produced hadron. This contribution
to the cross section is given by \cite{Boer:1997nt}
\begin{eqnarray}
& & \left. \frac{\D^5 \sigma^{(0)}_{\rm BM}}{\D x \, \D y \, 
\D z \, \D^2 \Vec P_T}
\right \vert_{\cos 2 \phi}
 =  \frac{4 \pi \alpha_{\rm em}^2 s}{Q^4} \, \sum_a
e_a^2 \, x (1 - y) \nonumber \\
& & \hspace{2cm} \times \,
 \int \D^2 \Vec k_T \, \int \D^2 \Vec p_T
\,\delta^2 (\Vec P_T - z \Vec k_T - \Vec p_T)
\nonumber \\
& & \hspace{2cm} \times \,
\frac{2 \, \Vec h \cdot \Vec k_T \,
\Vec h \cdot \Vec p_T - \Vec k_T \cdot \Vec p_T}{z  M M_h} \,
h_1^{\perp a} (x, k_T^2) \,
H_1^{\perp a} (z, p_T^2)\, \cos 2 \phi \,,
\label{cross3}
\end{eqnarray}
where $M$ is the mass of the nucleon and $M_h$ is the 
mass of the produced hadron. 
It should be noticed that this is a leading-twist contribution, not
suppressed by inverse powers of $Q$.

\begin{figure}
\begin{center}
\includegraphics[width=0.65\textwidth]
{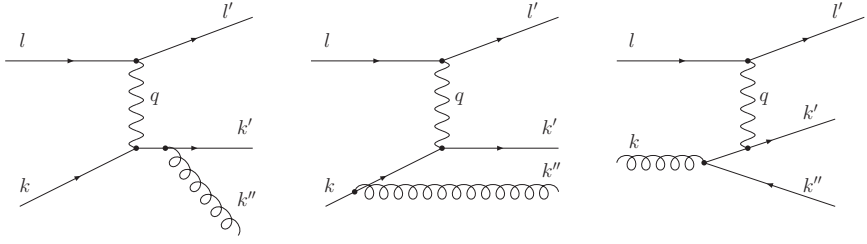}
\end{center}
\caption{\label{feynman} Feynman diagrams corresponding to $\ell q$
and $\ell g$ elementary scattering at first order in $\alpha_s$.}
\end{figure}

In our analysis we include both the non-perturbative 
and the perturbative contributions.
We consider now the contributions of order $\alpha_s$, 
following the approach of \cite{Chay:1992}. The
relevant partonic processes, shown in Fig.~\ref{feynman}, are those in
which the quark emits a hard gluon or those initiated by gluons:
\be
\gamma^* \,+\, q \to q \,+\, g \quad\quad
\gamma^* \,+\, q \to g \,+\, q \quad\quad
\gamma^* \,+\, g \to q \,+\, \bar{q}\,.
\ee
It is clear that, contrary to the lowest order process
$\gamma^* \,+\, q \to q$, the final parton can have a large transverse
momentum, even starting from a collinear configuration. Such a contribution
dominates the production of hadrons with large $P_T$ values.

We introduce the parton variables $\xp$ and $\zp$, defined
similarly to the hadronic variables $x$ and $z$,
\be
\xp = \frac {Q^2}{2k \cdot q}  = \frac{x}{\xi} \quad\quad
\zp = \frac {k \cdot k^\prime}{k \cdot q}  = \frac{z}{\zeta} \,,
\ee
where $k$ and $k^\prime$ are the four-momenta of the incident and fragmenting
partons, respectively. $\xi$ and $\zeta$ are the usual light-cone momentum
fractions, which, in the collinear configuration with massless partons, are
given by $k = \xi P$ and $P_h = \zeta k^\prime$. 
We denote by $\Vec \kappa_T$
the transverse momentum, with
respect to the $\gamma^*$ direction, of the final fragmenting parton,
$\Vec P_T = \zeta \Vec \kappa_T$.

The semi-inclusive DIS cross section, in the collinear QCD parton model, 
can be written in general as:
\bea
\label{eq:SIDIScrossoriginal}
\frac{\D^5\sigma}{\D x \, \D y \,\D z \,\D^2 \Vec P_T} &=&
\sum_{i,j} \int \D\xp \, \D\zp \, \D^2\Vec \kappa_T \, \D \xi \, \D\zeta \; 
\delta\left( x - \xi \xp \right) \delta\left( z - \zeta \zp \right)
\delta^2\left( \mathbf{P}_T - \zeta \Vec \kappa_T \right)
\nonumber \\
&& \times
f_1^i\left(\xi,Q^2\right)
\frac{d \hat{\sigma}_{ij}}{\D\xp \,\D y \,\D\zp \,\D^2\Vec \kappa_T}
D_1^j\left(\zeta,Q^2\right)\,.
\eea
%
%
To first order in $\alpha_s$ the partonic cross section is given by
\cite{Chay:1992}
\be
\frac{\D \hat{\sigma}_{ij}}{\D\xp \,\D y \,\D\zp \,\D^2\Vec \kappa_T} =
\frac{\alpha_{\rm em}^2e_q^2}{16\pi^2 Q^4} \,y \, L_{\mu\nu} M_{ij}^{\mu\nu}
\delta\left(\kappa_T^2 - \frac{\zp}{\xp}(1-\xp)(1-\zp)Q^2\right),
\ee
where $ij$ denote the initial and fragmenting partons, $ij = qq, qg, gq$.
Inserting the above expression into Eq. (\ref{eq:SIDIScrossoriginal}) yields,
for the $\mathcal{O}(\alpha_s)$ cross section \cite{Anselmino:2006rv}:
\be
\frac{\D^5\sigma^{(1)}}{\D x \, \D y \,\D z \,\D^2\bfP_T} =
\frac{\alpha_{\rm em}^2\,e_q^2}{16\pi^2}\, \frac{y}{Q^4}
\int_{x}^1 \frac{\D\xp}{\xp P_T^2 + z_h^2(1-\xp)Q^2}
\sum_{i,j} f_1^i \left(\frac{x}{\xp},Q^2\right)\, L_{\mu\nu}\,M_{ij}^{\mu\nu}
\,
D_1^j\left(z + \frac{\xp P_T^2}{z_h(1-\xp)Q^2},Q^2\right)\,
\label{LOxs}
\ee
with \cite{Chay:1992,Anselmino:2006rv}
\bea
L_{\mu\nu} M_{qq}^{\mu\nu} &=&\frac{64\pi \alpha_s}{3}Q^2
\frac{(l\cdot k)^2+(l^\prime \cdot k^\prime)^2 +
      (l^\prime \cdot k)^2 + (l \cdot k^\prime)^2}
     {(k\cdot k^{\prime\prime})(k^\prime \cdot k^{\prime\prime})} \nonumber \\
&=& \frac{64\pi\alpha_s}{3}\, \frac{Q^2}{y^2} \, \left\{
[ 1 + (1-y)^2]\;
\left[(1-\xp) (1-\zp) + \frac{1+ (\xp\zp)^2}{(1-\xp)(1-\zp)}\right]
\,+\,8\,\xp\zp\;(1-y)
\right.\nonumber \\
&& \left.-\; 4\,\sqrt{\frac{\xp\zp\;(1-y)}{(1-\xp)(1-\zp)}}\;(2-y)\;[\xp\zp
+ (1-\xp)(1-\zp)]\cos\phi \right.\nonumber \\
  &&  \left. + \;4\,\xp\zp\,(1-y)\,\cos 2\phi \frac{}{}\right\} \;,
\label{LOxs1}
\eea
\bea
L_{\mu\nu} M_{qg}^{\mu\nu}&=&\frac{64\pi \alpha_s}{3}Q^2
\frac{(l\cdot k)^2+(l^\prime \cdot k^{\prime\prime})^2 +
      (l^\prime \cdot k)^2 + (l \cdot k^{\prime\prime})^2}
     {(k\cdot k^\prime)(k^\prime k^{\prime\prime})} \nonumber \\
&=& \frac{64 \pi\alpha_s}{3}\,\frac{Q^2}{y^2}\,\left\{
[ 1 + (1-y)^2]\;
\left[(1-\xp)\zp + \frac{1+{\xp}^2(1-\zp)^2}{(1-\xp)\zp}\right]
\,+\,8\,\xp(1-y)(1-\zp)
\right.\nonumber \\
&& \left.+\; 4\,\sqrt{\frac{\xp(1-y)(1-\zp)}{(1-\xp)\zp}}\;(2-y)\;[\xp(1-\zp)
+ (1-\xp)\zp]\cos\phi \right.\nonumber \\
  &&  \left. + \;4\,\xp(1-y)(1-\zp)\,\cos 2\phi \frac{}{}\right\} \;,
\label{LOxs2}
\eea
\bea
L_{\mu\nu} M_{gq}^{\mu\nu}&=&\frac{64\pi \alpha_s}{3}Q^2
\frac{(l\cdot k^{\prime\prime})^2+(l^\prime \cdot k^\prime)^2 +
      (l^\prime \cdot k^{\prime\prime})^2 + (l \cdot k^\prime)^2}
     {(k\cdot k^\prime)( k\cdot k^{\prime\prime})} \nonumber \\
&=&  8\pi\alpha_s\;\frac{Q^2}{y^2}\left\{
[ 1 + (1-y)^2]\;[{\xp}^2+(1-\xp)^2]\;
\frac{{\zp}^2+(1-\zp)^2}{\zp(1-\zp)}
\,+\,16\,\xp (1-\xp) (1-y)
\right.\nonumber \\
&& \left.-\; 4\,\sqrt{\frac{\xp(1-\xp)(1-y)}{\zp(1-\zp)}}\;
(2-y)\;(1-2\xp)(1-2\zp)\,\cos\phi \right.\nonumber \\
&&  \left. + \;8\,\xp(1-\xp)(1-y)\,\cos 2\phi \frac{}{}\right\} \;,
\label{LOxs3}
\eea
where we have explicitely written the scalar products in terms of $\xp$, $y$,
$\zp$ and $\phi$. Notice the appearance of the $\cos\phi$ and
$\cos2\phi$ terms: $\phi$ is the azimuthal angle of the fragmenting
partons, which, in a collinear configuration, coincides with the azimuthal
angle of the detected final hadron. 
Since large values of $P_T$ cannot be generated by the modest amount of 
intrinsic
motion of quarks, we expect that Eq.(\ref{LOxs}) will dominantly describe
the cross sections for the lepto-production of hadrons with $P_T$ values
above $1$ GeV.

The  asymmetry determined experimentally is defined as
\begin{equation}
\langle \cos 2 \phi \rangle =
 \frac{\int \D \sigma \, \cos 2 \phi}{\int \D \sigma}\,.
\end{equation} 
The integrations are performed over the measured
ranges of $x, y, z$, with a lower cutoff $P_T^{\rm cut}$ on $P_T$, 
which represents the minimum value of $P_T$ of the detected charged particles.
Up to order $\alpha_s$ one has
\begin{equation} 
\langle \cos 2 \phi \rangle = 
\frac{\int \D \sigma^{(0)} \, \cos 2 \phi \; + \; 
\int \D \sigma^{(1)} \, \cos 2 \phi}{\int \D \sigma^{(0)}\; 
+ \;\int \D \sigma^{(1)}}\,,
\label{asymmetryfull}
\end{equation}
where $\sigma^{(0)}$ ($\sigma^{(1)}$) is the lowest order 
(first order) in $\alpha_s$  cross section. 
In the non perturbative region ($P_T \ll 1$ GeV) 
one expects $\sigma^{(0)} \gg \sigma^{(1)}$, 
thus to a very good approximation we have
\begin{equation}
\langle \cos 2 \phi \rangle \simeq \frac{\int \D \sigma^{(0)} \, 
\cos 2 \phi}{\int \D \sigma^{(0)}}\,,
\label{asymmetry0}
\end{equation}
Explicitly the numerator and the denominator are given by 
\begin{eqnarray}
& & 
\int \D \sigma^{(0)} \cos 2 \phi =
\frac{4 \pi \alpha_{\rm em}^2 s}{Q^4} \, 
\int \int \int \int \,
 \sum_a e_a^2 \,  x (1- y) \, \{ \mathcal{A} [f_1^a, D_1^a] +
\frac{1}{2} \, \mathcal{B} [h_1^{\perp a}, H_1^{\perp a}] \}\,, \\
& & \int \D \sigma^{(0)} =
\frac{2 \pi \alpha_{\rm em}^2 s}{Q^4} \, 
\int \int \int \int \,
\sum_a e_a^2 \, x   [1 +
(1 - y)^2]  \, \mathcal{C} [f_1^a,
D_1^a]\,, 
\end{eqnarray}
where
\begin{equation}
\int \int \int \int \equiv
\int_{P_T^{\rm cut}}^{P_{T}^{\rm max}} \D P_T \, P_T
\, \int_{x_1}^{x_2} \D x \,
\int_{y_1}^{y_2} \D y \, \int_{z_1}^{z_2} \D z \,
\end{equation}
and
($\chi$ is the angle between $\Vec P_T$
and $\Vec k_T$)
\begin{eqnarray}
\mathcal{A} [f_1^{a}, D_1^{a}] &\equiv&
 \int \D^2 \Vec k_T \, \int \D^2 \Vec p_T
\,\delta^2 (\Vec P_T - z \Vec k_T - \Vec p_T)
\nonumber \\
& &  \times \,
\frac{2 \, (\Vec k_T \cdot \Vec h)^2 -  k_T^2}{Q^2} \,
f_1^a (x,  k_T^2) \,
D_1^a (z,  p_T^2)\, \cos 2 \phi
\nonumber \\
& =&
\int_0^{\infty} \D k_T \, k_T \, \int_0^{2 \pi} \D \chi
\, \frac{2 \, k_T^2 \, \cos^2 \chi -  k_T^2}{Q^2}
\nonumber \\
& & \times \,  f_1^{a}(x,  k_T^2)
\, D_1^{a}(z, \vert \Vec P_T - z \Vec k_T \vert^2)\,,
\label{convol1}
\end{eqnarray}
\begin{eqnarray}
\mathcal{B} [h_1^{\perp a}, H_1^{\perp a}] &\equiv&
 \int \D^2 \Vec k_T \, \int \D^2 \Vec p_T
\,\delta^2 (\Vec P_T - z \Vec k_T - \Vec p_T)
\nonumber \\
& &  \times \,
\frac{2 \, \Vec h \cdot \Vec k_T \,
\Vec h \cdot \Vec p_T - \Vec k_T \cdot \Vec p_T}{z  M M_h} \,
h_1^{\perp a} (x,  k_T^2) \,
H_1^{\perp a} (z,  p_T^2)
\nonumber \\
& =&
\int_0^{\infty} \D k_T \, k_T \, \int_0^{2 \pi} \D \chi
\, \frac{ k_T^2 + (P_T/z)\, k_T
\, \cos \chi - 2 \,  k_T^2 \, \cos^2 \chi}{M M_h}
\nonumber \\
& & \times \, h_1^{\perp a}(x,  k_T^2)
\, H_1^{\perp a}(z, \vert \Vec P_T - z \Vec k_T \vert^2)\,,
\label{convol2}
\end{eqnarray}
\begin{eqnarray}
\mathcal{C} [f_1^{a}, D_1^{a}] &\equiv&
\int \D^2 \Vec k_T \,
\int \D^2 \Vec p_T \,
\delta^2 (\Vec P_T - z \Vec k_T - \Vec p_T)
\, f_1^a (x,  k_T^2) \,
D_1^a (z,  p_T^2)
\nonumber \\
& =&
\int_0^{\infty} \D k_T \, k_T \, \int_0^{2 \pi} \D \chi
\,  f_1^{a}(x,  k_T^2)
\, D_1^{a}(z, \vert \Vec P_T - z \Vec k_T \vert^2)\,.
\label{convol3}
\end{eqnarray}

\section{Parametrizations of the Boer--Mulders and Collins functions}

While the perturbative contribution to the $\cos 2 \phi$ asymmetry 
contain only standard distribution and fragmentation 
functions and therefore 
can be evaluated in a straightforward way, 
the non perturbative contributions involve the 
Boer--Mulders distribution, at present totally unknown, 
 and the Collins 
fragmentation function, for which we have some independent 
information coming from single-spin asymmetries in SIDIS and 
from $e^+ e^-$  data. 

Let us first consider the Boer--Mulders distribution. 
In order to estimate it, we resort  
to the impact--parameter approach
\cite{Burkardt:2003uw,Burkardt:2005hp,Burkardt:2005td,Diehl:2005jf},   
which establishes a general link  between 
the  anomalous tensor magnetic 
moment of quarks $\kappa_T^q$ and 
the transverse deformation of quark 
distributions in position space. This distortion  
translates, in a model-dependent way, 
into a single-spin asymmetry for 
transversely polarized quarks in unpolarized hadrons, 
and one finally finds a correlation between 
$h_1^{\perp q}$ and $\kappa_T^q$ (for explicit realizations
of this mechanism in spectator models and detailed discussions of its 
physical basis, see Ref.~\cite{Burkardt:2003je} and 
Ref.~\cite{Meissner:2007rx}),  
\[
h_1^{\perp q} \sim - \kappa_T^q \,.
\] 
Since a similar connection exists 
between the Sivers function $f_{1T}^{\perp q}$ and the anomalous 
magnetic moment $\kappa^q$, 
\[
f_{1T}^{\perp q} \sim - \kappa^q \,, 
\] 
 the sizes 
of $h_1^{\perp}$ and $f_{1T}^{\perp}$ 
are expected \cite{Burkardt:2005hp,Burkardt:2005td} to be roughly 
alike, up to a scale factor $\kappa_T^q / \kappa^{q}$:  
\be
h_1^{\perp q}(x, k_T^2) = \frac{\kappa_T^q}{\kappa^{q}} 
f_{1T}^{\perp q}(x, k_T^2)
\label{burkardt}
\ee 
The contributions of $u$ and $d$ quarks to the anomalous magnetic moment 
of the proton ($\kappa^p$) and of the neutron ($\kappa^n$)
can be extracted from the experimental values $\kappa^p = 1.79$ 
and $\kappa^n = -1.91$ \cite{Yao:2006px} by means of $\kappa^{p,n} 
= \sum_q e_q \kappa^q$
(neglecting the $s$ quark contribution), and are given by
$\kappa^{u} \simeq 1.67$
and $\kappa^{d} \simeq -2.03$: note the opposite sign,  
which explains the experimentally observed fact that 
$f_{1T}^{\perp u} <0$ while  $f_{1T}^{\perp d} >0$. 
The flavor contributions to the anomalous tensor
magnetic moment have been  
 estimated by a lattice calculation in Ref.~\cite{Gockeler:2006zu} 
and are found to be $ \kappa_T^{u} \simeq 3$,  
$\kappa_T^{d} \simeq 1.9$. Thus, at variance 
with $f_{1T}^{\perp}$, we expect the $u$ and 
the $d$ components of $h_1^{\perp}$ to have the same 
sign, and in particular to be both negative (this qualitative expectation is 
also supported by large--$N_c$ arguments \cite{Pobylitsa:2003} 
and by various model calculations 
\cite{Yuan:2003wk,Pasquini:2005dk,Burkardt:2007xm,Gamberg:2007wm}).   
Inserting the values of $\kappa^q$ and 
 $\kappa_T^q$ in (\ref{burkardt}), one has
\be
h_1^{\perp u}\simeq  1.80 \, 
f_{1T}^{\perp u}\,, \;\;\;
h_1^{\perp d} = - 0.94 \, 
f_{1T}^{\perp d}
\label{burkardt2}
\ee 
Thus the $u$ component of $h_1^{\perp}$ is about twice as large 
as the corresponding component of $f_{1T}^{\perp}$, while the 
$d$ components of $h_1^{\perp}$ and $f_{1T}^{\perp}$ have 
approximately the same 
magnitude and opposite sign. 
To parametrize the Boer--Mulders function we use the 
Ansatz (\ref{burkardt2}) and 
get the Sivers function from a fit of 
single--spin asymmetry 
data \cite{Anselmino:2005ea}. The parametrization of 
Ref.~\cite{Anselmino:2005ea} 
for $f_{1T}^{\perp}$ is 
\be
f_{1T}^{\perp q} (x, k_T^2)=  \rho_q(x) \, \eta(k_T) \, 
f_1^q(x, k_T^2) \, , \label{sivfac}
\ee
where
\bea
&&\rho_q(x) =  A_q \, x^{a_q}(1-x)^{b_q} \,
\frac{(a_q+b_q)^{(a_q+b_q)}}{a_q^{a_q} b_q^{b_q}}\; , 
\label{siversx} \\
&&\eta (k_T) = \frac{2 M M_0}{k_T^2+ M_0^2}\; \cdot
\label{siverskt}
\eea
Here $A_q$, $a_q$, $b_q$ and $M_0$ are free parameters, 
$f_1^q (x,k_T)$ is the $k_T$-dependent unpolarized distribution function, 
which we assume to have a Gaussian behavior in $k_T$: 
\be
f_1^q (x, k_T^2) = f_1^q (x) \frac{\E^{-{ k_T^2}/
{\langle  k_T^2 \rangle}}}{\pi \langle  k_T^2 \rangle }\,. 
\ee
Notice that $f_{1T}^{\perp}$, being a quark spin asymmetry, 
must satisfy a positivity bound. This bound and 
the valence number sum rules are automatically fulfilled 
by the parametrization of Ref.~\cite{Anselmino:2005ea}. 
The average value of the intrinsic transverse momentum 
of quarks is taken from the SIDIS fit of Ref.~\cite{Anselmino:2005nn}, 
and is given by  
$\langle k_T^2 \rangle = 0.25$ GeV$^2$. This value is assumed 
to be constant and flavor-independent. It is worth recalling 
that the Gaussian behavior of 
$k_T$--dependent distribution functions 
is supported by a recent lattice study \cite{Musch:2007}, 
which finds a root mean squared transverse momentum 
very close to the one determined in Ref.~\cite{Anselmino:2005nn}
and used here. 
The fitted parameters in 
Eqs.~(\ref{siversx},\ref{siverskt})
are given in Table~\ref{fitpar} \cite{Anselmino:2005ea}.
\begin{table}[t]
\begin{center}
\begin{tabular}{|ll|ll|}
\hline
$A_{u}$  = & $-0.32  \pm  0.11$ & $A_{d}$ = & $1.00  \pm  0.12$ \\
$a_{u}$  = & $0.29  \pm  0.35$ & $a_{d}$ = & $ 1.16  \pm  0.47$ \\
$b_{u}$  = & $0.53  \pm  3.58$ & $b_{d}$ = & $ 3.77  \pm  2.59$ \\
\hline
$M_0^2$ = & $0.32 \pm 0.25 \; {\rm GeV}^2$ &  
&  \\
\hline
\end{tabular}
\end{center}
\caption{\small Best fit values of the parameters of the Sivers function.    
\label{fitpar}}
\end{table}

Let us now turn to the Collins function. 
We will distinguish the favored and the 
unfavored fragmentation functions according to 
the following general relations
\bea
&& D_{\pi^+/u} = D_{\pi^+/\bar d} = D_{\pi^-/d} = D_{\pi^-/\bar u}
\equiv D_{\rm fav} \label{fav} \\
&& D_{\pi^+/d} = D_{\pi^+/\bar u} = D_{\pi^-/u} = D_{\pi^-/\bar d}
= D_{\pi^\pm/s} =  D_{\pi^\pm/\bar s} \equiv D_{\rm unf}, \label{unf}
\eea
For the Collins function we use the parametrization 
of \cite{Anselmino:2007fs}, based on a combined 
analysis of SIDIS and $e^+ e^-$ data: 
\begin{equation} 
H_1^{\perp q} (z, p_T^2) = 
\rho_q^C(z) \, \eta^C(p_T) \, D_1^q (z, p_T^2)\,, 
\label{coll-funct}
\end{equation} 
with
\bea
&&\rho_q^C(z)= A_q^C \, z^{\gamma} (1-z)^{\delta} \,
\frac{(\gamma + \delta)^{(\gamma +\delta)}}
{\gamma^{\gamma} \delta^{\delta}} \label{NC}\\
&&
\eta^C(p_T)=
\sqrt{2 e}\,\frac{z M_h}{M_C} \, 
\E^{-{p_T^2}/{M_C^2}}\,,\label{h-funct}
\eea
We let the coefficients $A_q^C$ to be flavor dependent
($q = u,d)$, while all the exponents $\gamma, \delta$ and the
dimensional parameter $M_C$ are taken to be flavour independent.
The  parameterization is devised in such a way that the Collins function 
satisfies the positivity bound (remember that 
$H_1^{\perp}$ is essentially a transverse momentum 
asymmetry).  In Eq.~(\ref{coll-funct}) $D_1 (z, p_T^2)$
is the $p_T$--dependent unpolarized fragmentation, that we take to be 
given by 
\begin{equation} 
D_1^q(z, p_T^2) = D_1^q (z) \, \frac{\E^{- p_T^2/\langle p_T^2 \rangle}}{\pi 
\langle p_T^2 \rangle} 
\end{equation} 
assuming the usual gaussian behavior in $p_T$. 
Again, the average value of $p_T^2$ is taken from the fit 
of Ref.~\cite{Anselmino:2005nn} to the azimuthal 
dependence of the unpolarized 
SIDIS cross section: $\langle p_T^2 \rangle = 0.20$ GeV$^2$.  

The values of the parameters as determined in the fit 
of Ref.~\cite{Anselmino:2007fs} are listed in Table~\ref{fitpar1}.

Finally, we need the ordinary unpolarized distribution 
and fragmentation functions, $f_1(x)$ and $D_1(z)$, appearing 
both in the non-perturbative and in the perturbative 
contributions. They are taken from the GRV98 \cite{Gluck:1998xa}
and the Kretzer \cite{Kretzer:2000yf} parametrizations, 
respectively.

\begin{table}[t]
\begin{center}
\begin{tabular}{|l|llllll|}
\hline
~&~&~&~&~&~&~\\
~~Collins &
~~$A_{fav}^C$  &=& $0.41  \pm  0.91$ & $A_{unf}^C$   &=& $-1.00  \pm  0.96$ \\
~~fragmentation~~ &
~~$\gamma$  &=& $1.04  \pm  0.38$ & $\delta$   &=& $0.13  \pm  0.25$    \\
~~function &
~~$\langle p_T^2 \rangle $  &=& 0.2 GeV$^2$ & $M_C^2$  
&=&  $(0.71 \pm 0.65)$ GeV$^2$ ~\\
~&~&~&~&~&~&~\\
\hline
\end{tabular}
\end{center}
\caption{Best values of the favored and unfavored Collins
fragmentation functions. 
\label{fitpar1}}
\end{table}
%

\section{Results and predictions for $\langle \cos 2 \phi \rangle$}

\begin{figure}[t]
\includegraphics[width=0.5\textwidth,bb= 10 140 540 660,angle=-90]
{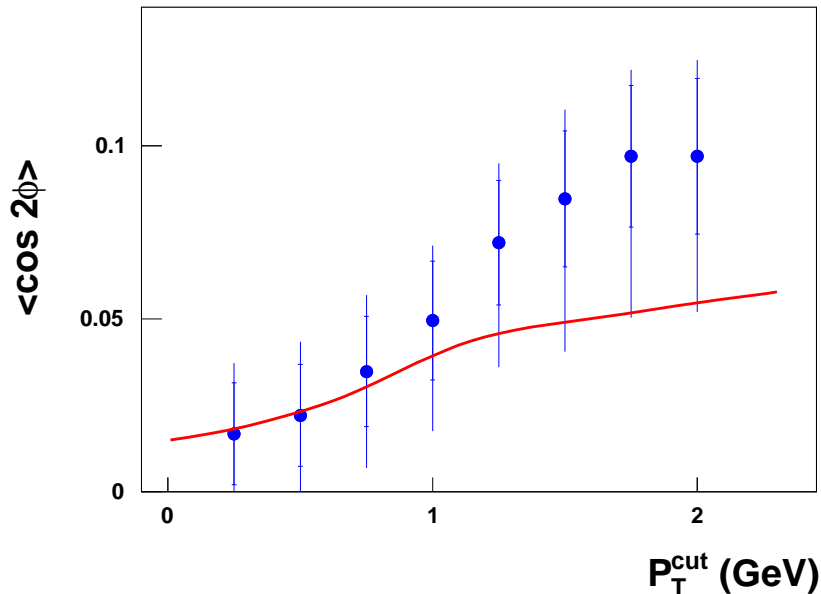}
\caption{\label{fig:zeus.ps}
Our prediction for $ \langle \cos 2\phi \rangle$ in 
charged pion production at ZEUS,  
compared with the data. The asymmetry is completely 
dominated by the perturbative contribution. 
}
\end{figure}

To start with, we compare our results 
with the available large-$Q^2$ data 
from the ZEUS collaboration (positron-proton collisions at
$300$ GeV) \cite{Breitweg:2000qh}. The integrations are performed 
over the following experimental ranges: 
\begin{equation}
 0.01 < x < 0.1 \;,\;\;\;0.2 < y < 0.8 \;,\;\;\;
0.2 < z < 1.0\;. 
\end{equation}
As shown in Fig.~\ref{fig:zeus.ps}, we find quite a  
good agreement with the data. Notice that 
the average $Q^2$ value, $\langle Q^2 \rangle \simeq 750$ GeV$^2$, 
is such that the asymmetry is completely dominated by 
the perturbative contribution.

In order to highlight the effect of the non perturbative contributions
(Boer--Mulders and higher twist) one has to probe the 
kinematical region corresponding to $P_T < 1$ GeV 
and $Q^2$ of order of few GeV$^2$, where 
the gluon emission is quite irrelevant.  
%
Such a testing ground is investigated at the
 HERMES, COMPASS and JLAb facilities. We now turn to our predictions 
for these experiments.

\begin{figure}[t]
\includegraphics[width=0.5\textwidth,bb= 10 140 540 660,angle=-90]
{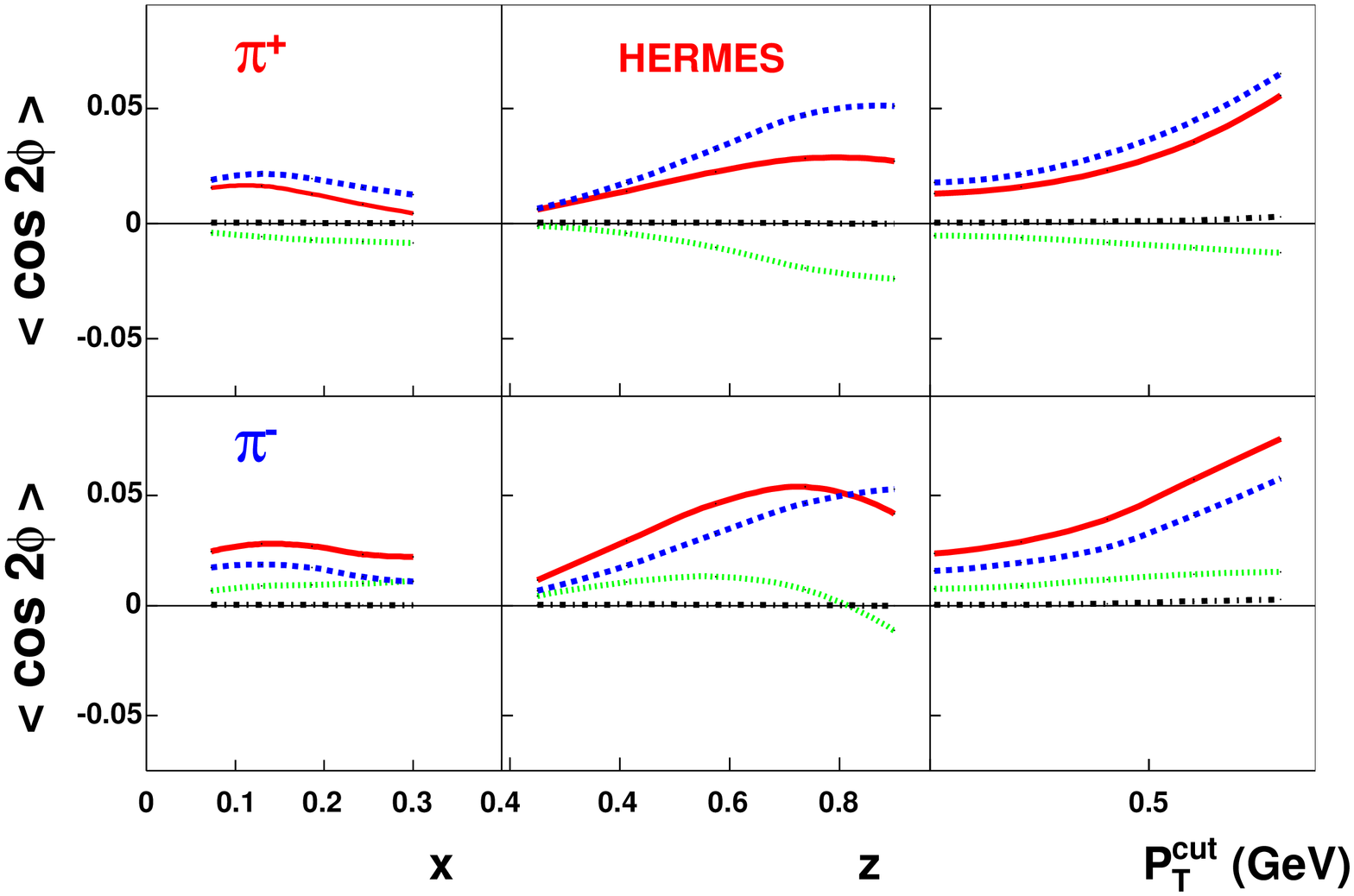}
\caption{\label{fig:hermes.ps}
Our prediction for the $\cos 2\phi$ asymmetry at HERMES.
The dot--dashed line is  
the ${\cal O}(\alpha_s)$ QCD contribution,
the dotted line is the Boer-Mulder contribution,  
the dashed line is the higher--twist Cahn contribution.  
The continuous line is the resulting 
asymmetry taking all contributions into account.}
\end{figure}

\begin{figure}[t]
\includegraphics[width=0.5\textwidth,bb= 10 140 540 660,angle=-90]
{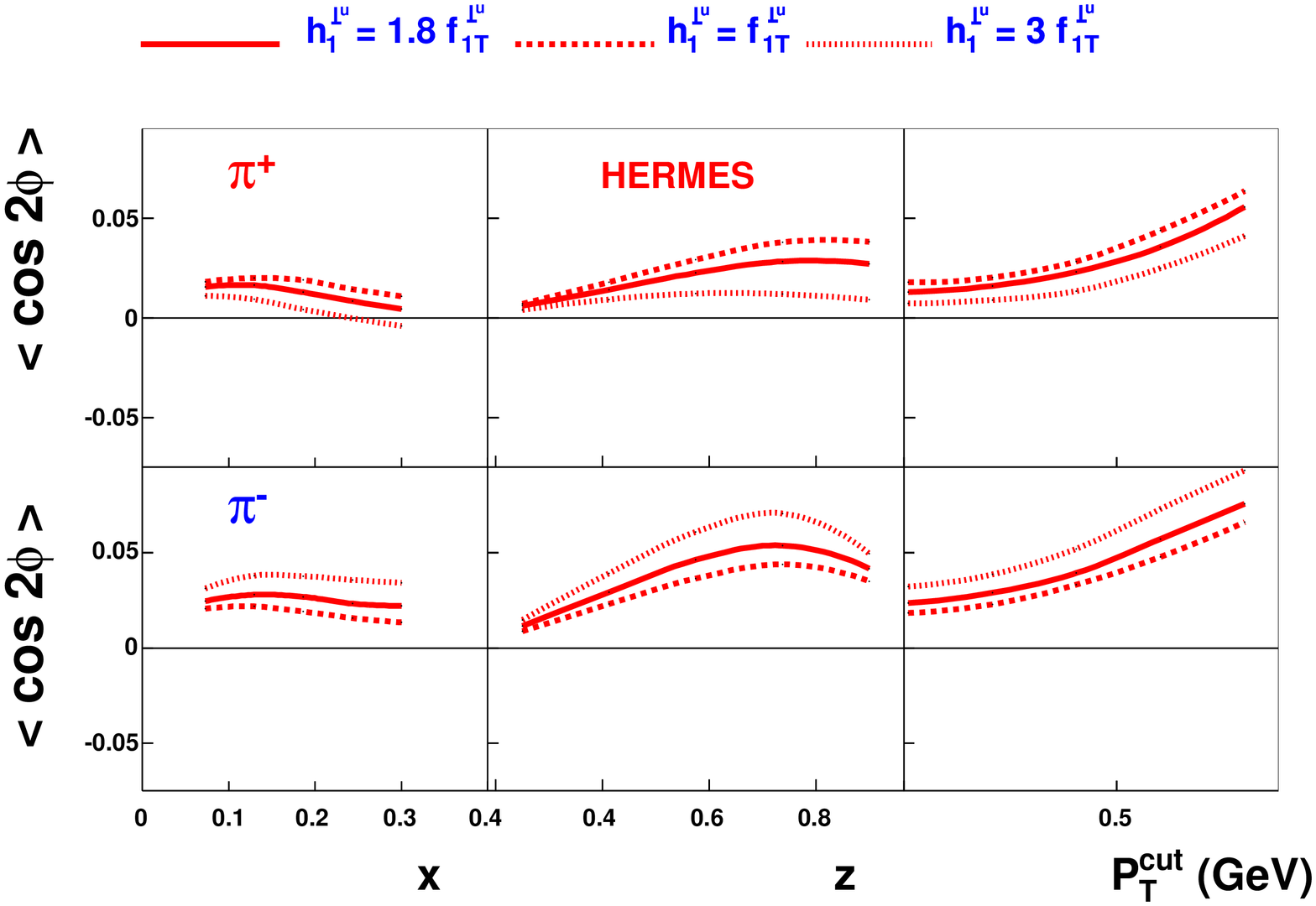}
\caption{\label{fig:hermes_factor.ps}
Our prediction for the $\cos 2\phi$ asymmetry at HERMES, 
with three different assumptions for $h_1^{\perp u}$. The solid 
line corresponds to the Ansatz adopted here. 
}
\end{figure}

In Fig.~\ref{fig:hermes.ps}, we plot  $\langle
\cos2\phi \rangle$ for $\pi^+$ and $\pi^-$
production at HERMES, as a function of one 
variable at a time, $x$,
$z$ and $P_T^{\rm cut}$; the integration over the unobserved variables has been
performed over the measured ranges of the HERMES experiment, 
\bea \nonumber && Q^2 > 1 \; {\rm GeV}^2\,, \quad W^2 > 10 \; 
{\rm GeV}^2 \,, 
\quad P_T > 0.05 \; {\rm GeV} \\
\label{hermutcuts} && 0.023 < x < 0.4\,, \quad 0.2 < z < 0.7 \,, \quad
0.1 < y < 0.85 \nonumber \\ && 2 < E_h < 15 \> {\rm GeV} \>. \eea
In these kinematical regions the cross section is dominated by 
$\sigma^{(0)}$. An interesting feature of the asymmetry is that 
the Boer-Mulders contributions 
to $\pi^+$ and $\pi^-$ production are opposite in sign. In fact, we have
\bea
\langle \cos 2 \phi \rangle^{\pi^+}_{\rm BM} \sim e_u^2 \, 
h_1^{\perp u}(x) \, H_1^{\perp {\rm fav}}(z) + e_d^2 \, h_1^{\perp d}(x) \, 
H_1^{\perp {\rm unf}}(z) \, ,\nonumber \\
\langle \cos 2 \phi \rangle^{\pi^-}_{\rm BM} \sim e_u^2 \, 
h_1^{\perp u}(x)\, H_1^{\perp {\rm unf}}(z) + e_d^2 \, h_1^{\perp d}(x) \, 
H_1^{\perp {\rm fav}}(z) \, ,
\eea
and, as far as $H_1^{\perp {\rm unf}}(z) \simeq - H_1^{\perp {\rm fav}}(z)$ 
\cite{Anselmino:2007fs}, one gets different signs for the 
Boer-Mulders effect for positive and negative pions. 
Another important finding is the quantitative relevance of the 
higher--twist (Cahn) component of $\langle \cos 2 \phi \rangle$.  
This contribution, which is positive,
 is the same for $\pi^+$ and $\pi^-$, if
 the $k_T$--dependence of the quark distributions 
is flavor--independent. 
Thus the asymmetry resulting from  the combination of the 
Boer--Mulders and Cahn contributions turns out to be 
larger for $\pi^-$ than for $\pi^+$. We conclude that 
a difference between $\langle \cos 2 \phi \rangle^{\pi^-}$ 
and $\langle \cos 2 \phi \rangle^{\pi^+}$ is a clear 
signature of the Boer--Mulders effect. This 
is a definite prediction, to be checked experimentally. 
Moreover, the bigger in 
magnitude $h_1^{\perp}$,   
the more pronounced is the difference between the $\pi^-$ and the $\pi^+$ 
asymmetry. To illustrate this, we show in  Fig.~\ref{fig:hermes_factor.ps}
our predictions for $\langle \cos 2 \phi \rangle$ with 
three different choices of $h_1^{\perp}$: one corresponding to 
the Ansatz (\ref{burkardt2}), the other two corresponding 
to a smaller and to a larger (in magnitude) $h_1^{\perp}$.

\begin{figure}[t]
\includegraphics[width=0.5\textwidth,bb= 10 140 540 660,angle=-90]
{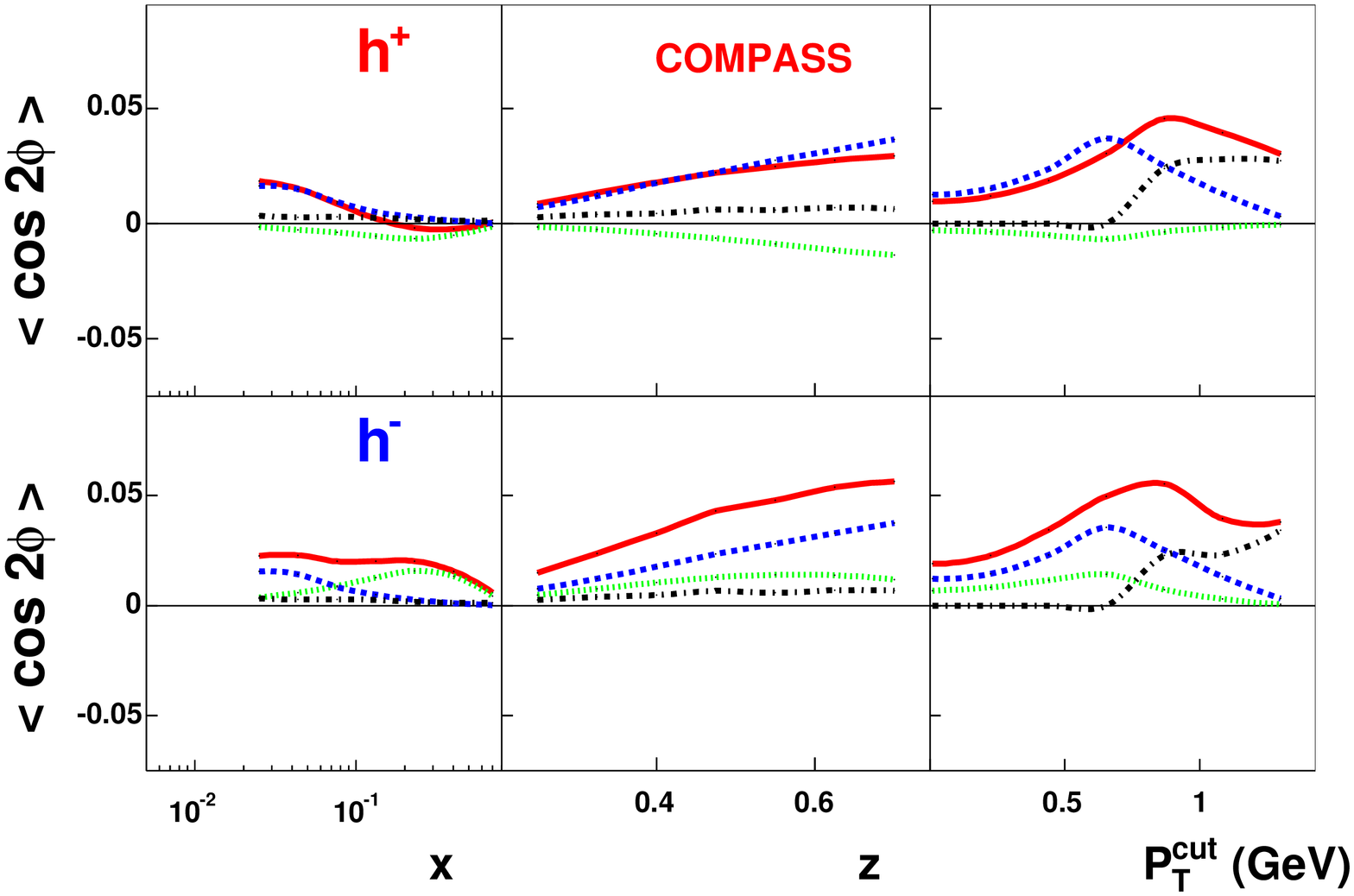}
\caption{\label{fig:compass.ps}
Our prediction for the $\cos 2\phi$ asymmetry at COMPASS, with a 
deuteron target.
The line labels are the same as in Fig.~\ref{fig:hermes.ps}}
\end{figure}

Fig.~\ref{fig:compass.ps} shows our predictions for 
$\langle \cos 2\phi_h\rangle$ at COMPASS 
with a deuteron target. The experimental cuts are
\bea 
&& Q^2 \geq 1 \; {\rm GeV}^2\,, \quad W^2 \geq 25 \; 
{\rm GeV}^2 \,, \\ \nonumber
 && 0.2 \leq z \leq 1 \,, \quad
0.1 \leq y \leq 0.9 \,, \label{Compass} \\ 
  && E_h \leq 15 \, {\rm GeV}\,.  
\nonumber
 \eea 
We neglect nuclear corrections and
use isospin symmetry to relate the distribution functions of 
the neutron to those of the proton. 
For the Boer--Mulders contribution we have
\bea
\langle \cos 2 \phi \rangle^{\pi^+/ D}_{BM} 
\sim (h_1^{\perp u}(x) + h_1^{\perp d}(x))\,
(  e_u^2 \, H_1^{\perp fav}(z) + e_d^2 \, H_1^{\perp unf}(z) ) \, ,
\nonumber \\
\langle \cos 2 \phi \rangle^{\pi^-/ D}_{BM} \sim 
(h_1^{\perp u}(x) + h_1^{\perp d}(x))\,(  e_u^2 \, 
H_1^{\perp unf}(z) + e_d^2 \, H_1^{\perp fav}(z) ) \,. 
\eea
Since $h_1^{\perp u}$ has the same sign as $h_1^{\perp d}$ 
the deuteron target tends to exalt the Boer--Mulders effect. The    
opposite happens for the Sivers effect 
in transversely polarized SIDIS, which is suppressed 
for a deuteron target since 
 $f_{1T}^{\perp u}$ and $f_{1T}^{\perp d}$, 
having different sign,  partly cancel each other.  
Notice also that  the perturbative contribution 
to the asymmetry at COMPASS becomes 
non negligible for $P_T > 1$ GeV.

\begin{figure}[t]
\includegraphics[width=0.5\textwidth,bb= 10 140 540 660,angle=-90]
{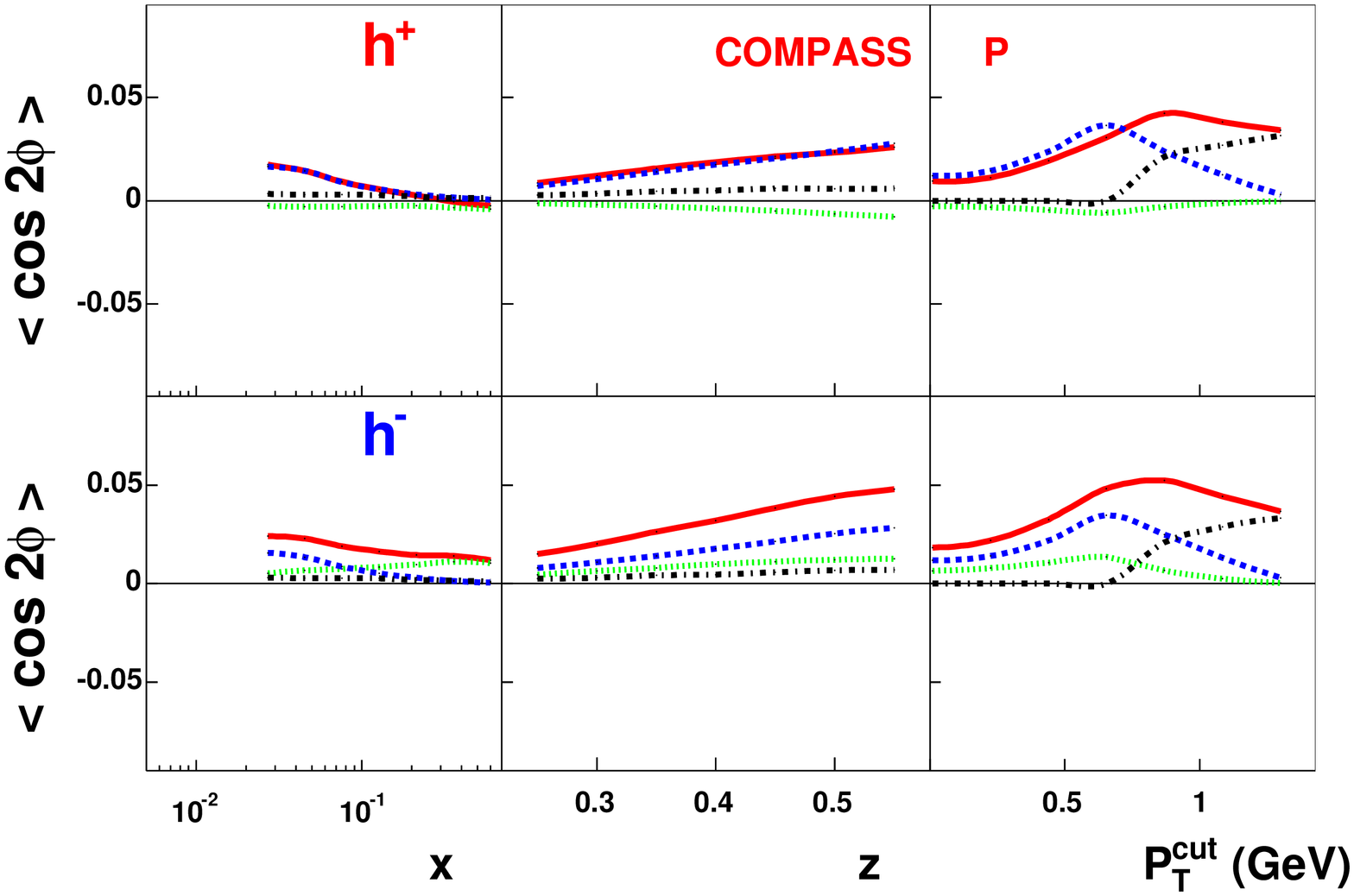}
\caption{\label{fig:compass_proton.ps}
Our prediction for the $\cos 2\phi$ asymmetry at COMPASS, with a  
proton target.The line labels are the same 
as in Fig.~\ref{fig:hermes.ps}}
\end{figure}

For completeness, in Fig.~\ref{fig:compass_proton.ps} we present
our predictions for the COMPASS experiment operating with a proton target.

\begin{figure}[t]
\includegraphics[width=0.5\textwidth,bb= 10 140 540 660,angle=-90]
{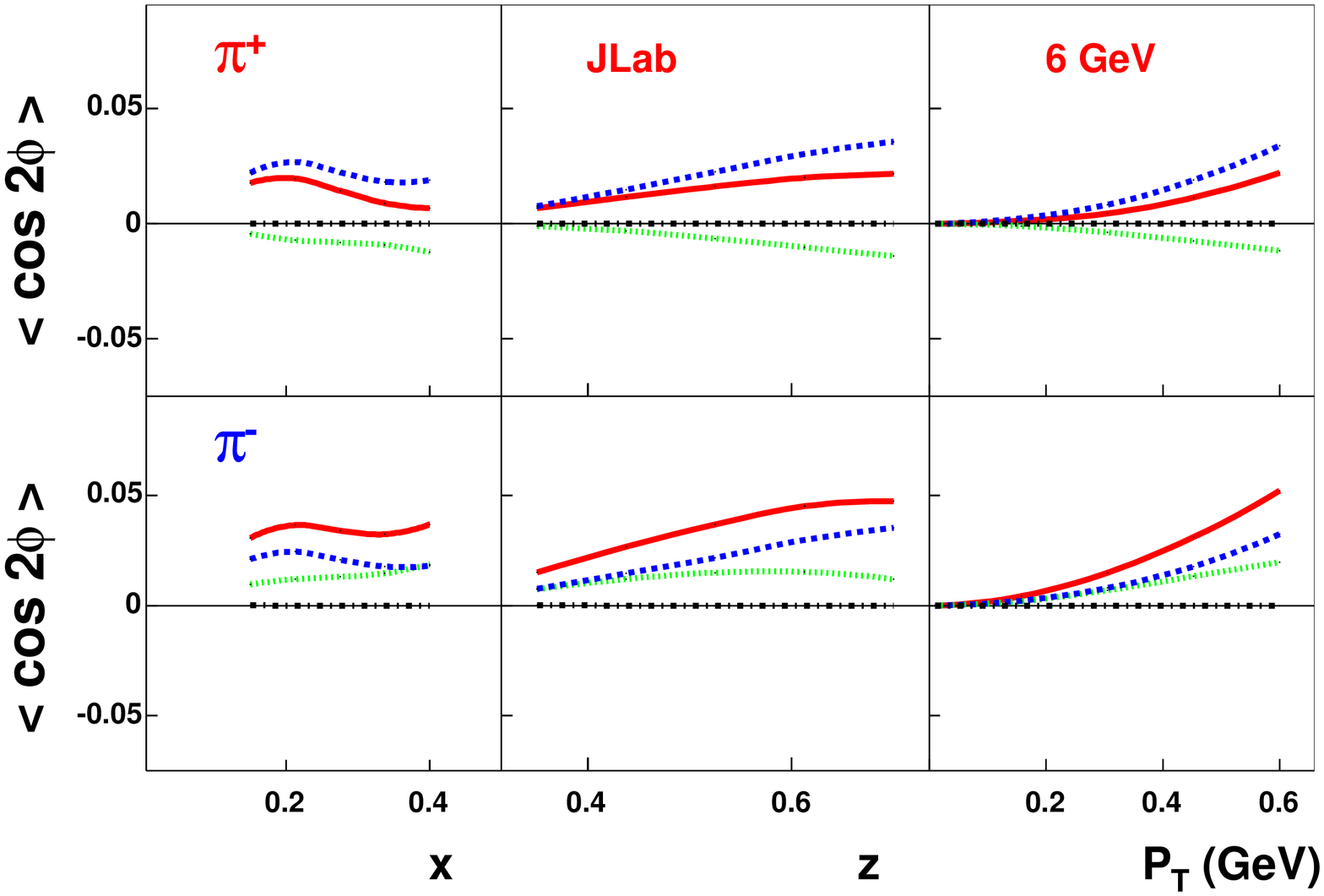}
\caption{\label{fig:jlab6.ps}
Our prediction for $\langle \cos 2\phi\rangle$ at JLab,  
with an incident beam energy of 6 GeV operating and a proton target. 
The line labels are the same as in Fig.~\ref{fig:hermes.ps}}
\end{figure}

\begin{figure}[t]
\includegraphics[width=0.5\textwidth,bb= 10 140 540 660,angle=-90]
{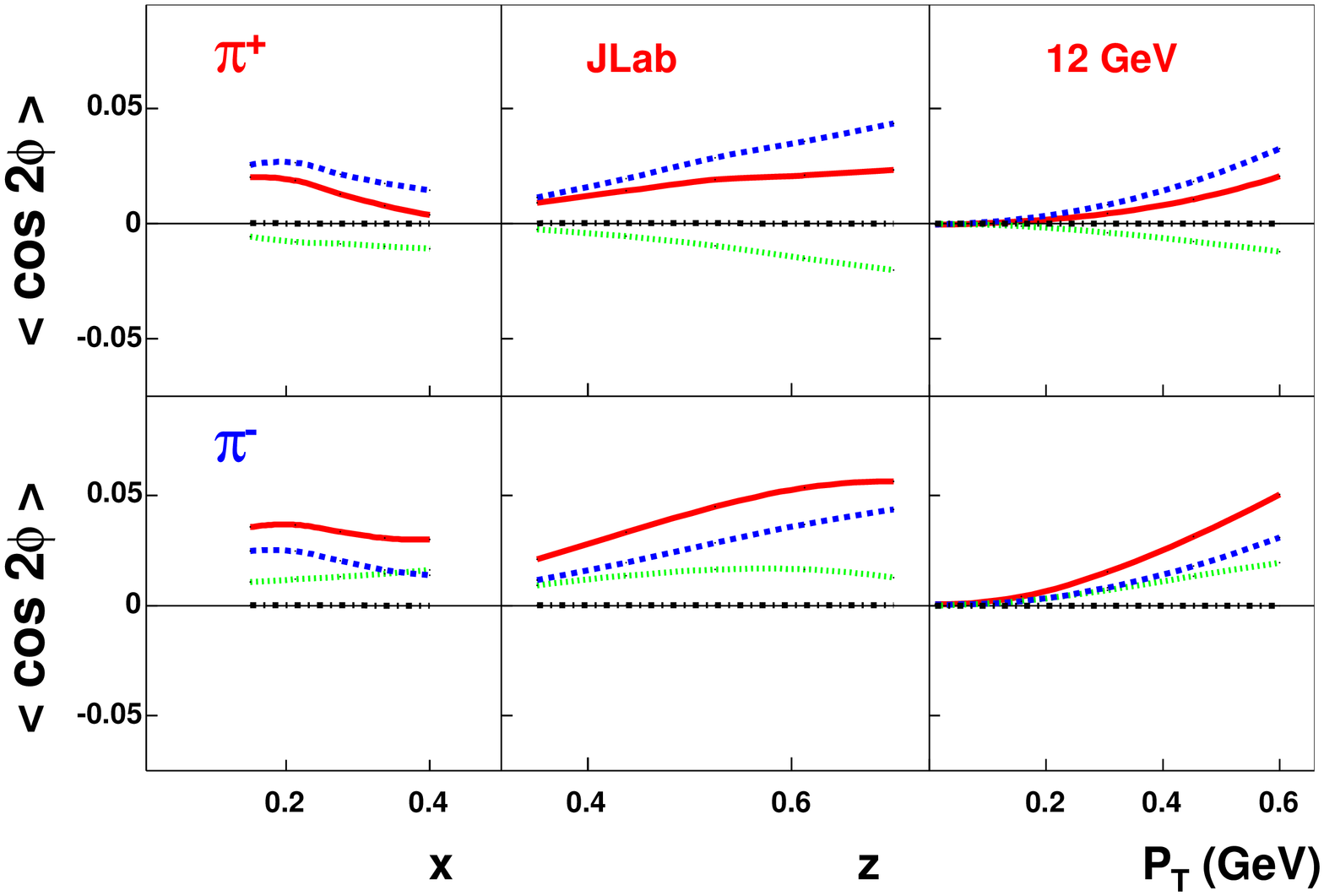}
\caption{\label{fig:jlab12.ps}
Our prediction for $\langle \cos 2\phi \rangle$ at JLab,  
with an  incident beam energy of 12 GeV operating and a proton target. 
The line labels are the same as in Fig.~\ref{fig:hermes.ps}}
\end{figure}

\begin{figure}[t]
\includegraphics[width=0.5\textwidth,bb= 10 140 540 660,angle=-90]
{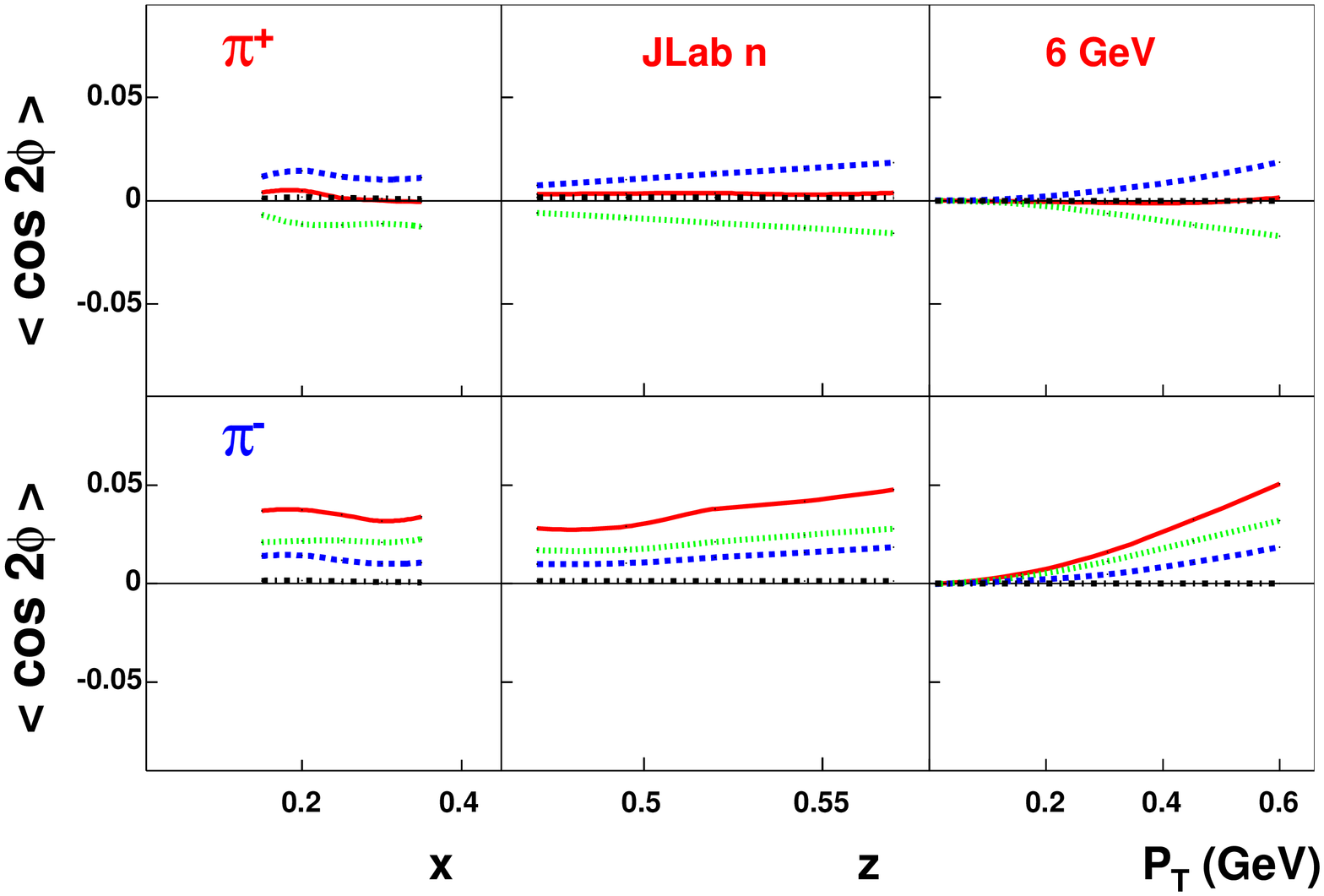}
\caption{\label{fig:jlab6_neutron.ps}
Our prediction for $\langle \cos 2\phi \rangle$ at JLab,  
with an incident beam energy of 6 GeV and  a neutron target. 
The line labels are the same as in Fig.~\ref{fig:hermes.ps}}
\end{figure}

\begin{figure}[t]
\includegraphics[width=0.5\textwidth,bb= 10 140 540 660,angle=-90]
{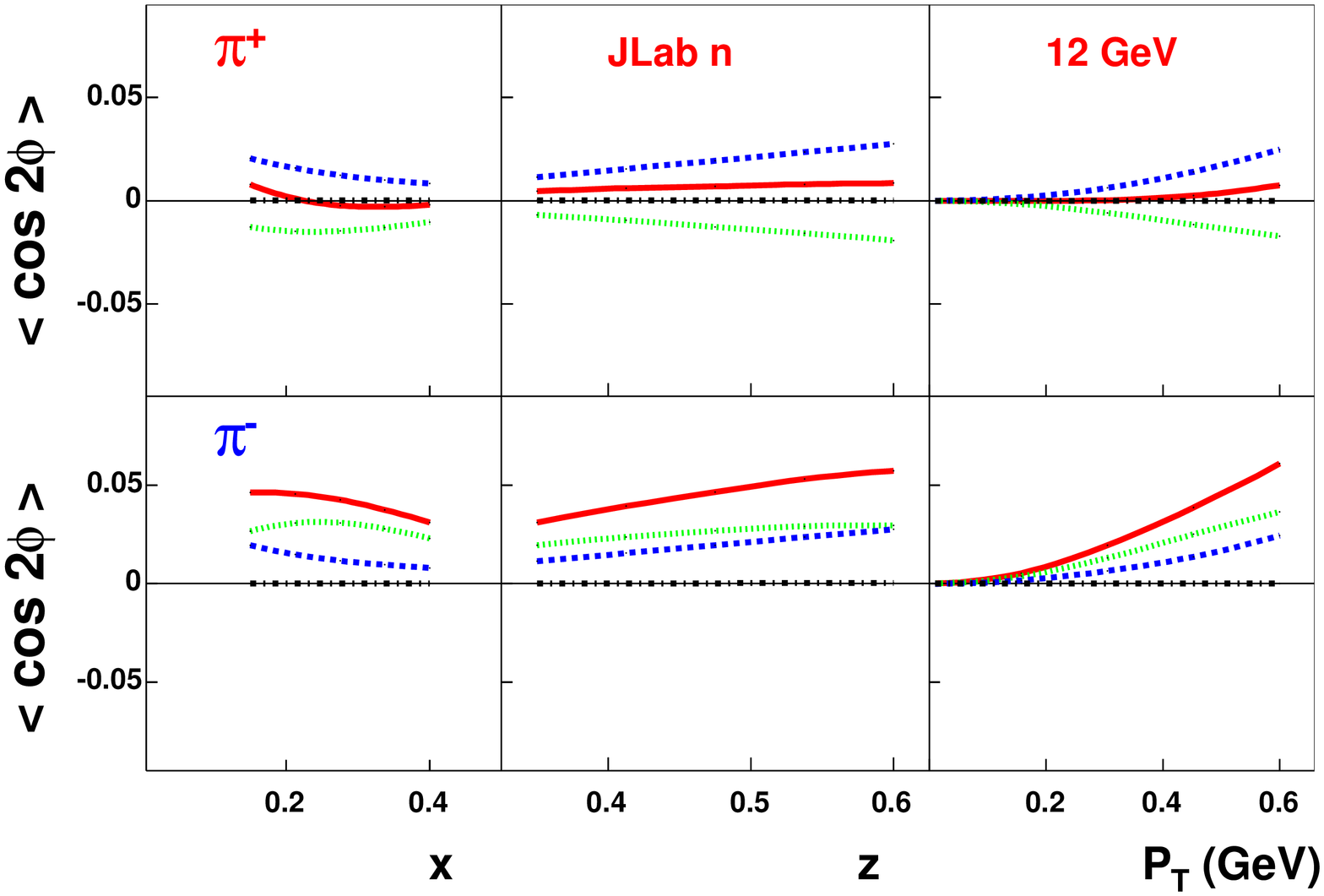}
\caption{\label{fig:jlab12_neutron.ps}
Our prediction for $\langle \cos 2\phi \rangle$ asymmetry at JLab,  
with an incident beam energy of 12 GeV and a neutron target. 
The line labels are the same as in Fig.~\ref{fig:hermes.ps}}
\end{figure}

Finally, JLab collects data in the collisions of $6$ and $12$
GeV electrons from proton and neutron targets. 
The experimental cuts for JLab operating with a 
proton target and a $6$ GeV beam are the following
\bea 
 && Q^2 \geq 1 \; {\rm GeV}^2\,, \quad W^2 \geq 4 \; 
{\rm GeV}^2 \,, 
\quad 0.02 \leq P_T \leq 1 \; {\rm GeV} \nonumber \\
 && 0.1 \leq x \leq 0.6\,, \quad 0.4 \leq z \leq 0.7 \,, \quad
0.4 \leq y \leq 0.85 \label{JLab-6} \\
  && 1 \leq E_h \leq 4 \, {\rm GeV}\,,  
\nonumber
 \eea
whereas for a beam energy of $12$ GeV they are
\bea 
 && Q^2 \geq 1 \; {\rm GeV}^2\,, \quad W^2 \geq 4 \; 
{\rm GeV}^2 \,, 
\quad 0.02 \leq P_T \leq 1.4 \; {\rm GeV} \nonumber \\
 && 0.05 \leq x \leq 0.7\,, \quad 0.4 \leq z \leq 0.7 \,, \quad
0.2 \leq y \leq 0.85 \label{JLab-12} \\
  && 1 \leq E_h \leq 7 \, {\rm GeV}\,.  
\nonumber
 \eea
For the neutron target with the $6$ GeV beam the cuts are:
\bea 
 && 1.3 \leq Q^2 \leq 3.1 \; {\rm GeV}^2\,, \quad 5.4 \leq W^2 \leq 9.3 \; 
{\rm GeV}^2 \,, \nonumber \\
 && 0.13 \leq x \leq 0.4\,, \quad 0.46 \leq z \leq 0.59 \,, \quad
0.68 \leq y \leq 0.86 \label{JLab-6n} \\
  && 2.385 \leq E_h \leq 2.404 \, {\rm GeV}\,,  
\nonumber
 \eea
whereas for the incident beam energy of $12$ GeV they are:
\bea 
 &&  Q^2 \geq 1 \; {\rm GeV}^2\,, \quad  W^2 \geq 2.3 \; 
{\rm GeV}^2 \,, \nonumber \\
 && 0.05 \leq x \leq 0.55\,, \quad 0.3 \leq z \leq 0.7 \,, \quad
0.34 \leq y \leq 0.9 \label{JLab-12n} 
 \eea
Our results for $\langle \cos 2\phi_h\rangle$ are shown in
Figs.~\ref{fig:jlab6.ps},~\ref{fig:jlab12.ps},~\ref{fig:jlab6_neutron.ps},
 and~\ref{fig:jlab12_neutron.ps}. 
Notice that unlike the predictions 
for HERMES and COMPASS, where the dependence on $P_T^{\rm cut}$ is presented,  
for JLab we show the dependence
on $P_T$ (thus $\langle \cos 2 \phi \rangle$ vanishes when $P_T\rightarrow 0$).
As one can see, 
the JLab measurements are insensitive to the
perturbative QCD corrections and completely dominated 
by $\mathcal{O}(\alpha_s^0)$ effects. Again, due to 
the Boer--Mulders contribution, the $\pi^-$ asymmetry is 
larger than the $\pi^+$ asymmetry. In particular, in 
the case of a neutron target, the Boer--Mulders effect and 
the Cahn higher twist effect combine to yield a vanishing 
$\pi^+$ asymmetry and a 4-5 \%  $\pi^-$ asymmetry.  

\begin{figure}[t]
\includegraphics[width=0.5\textwidth,bb= 10 140 540 660,angle=-90]
{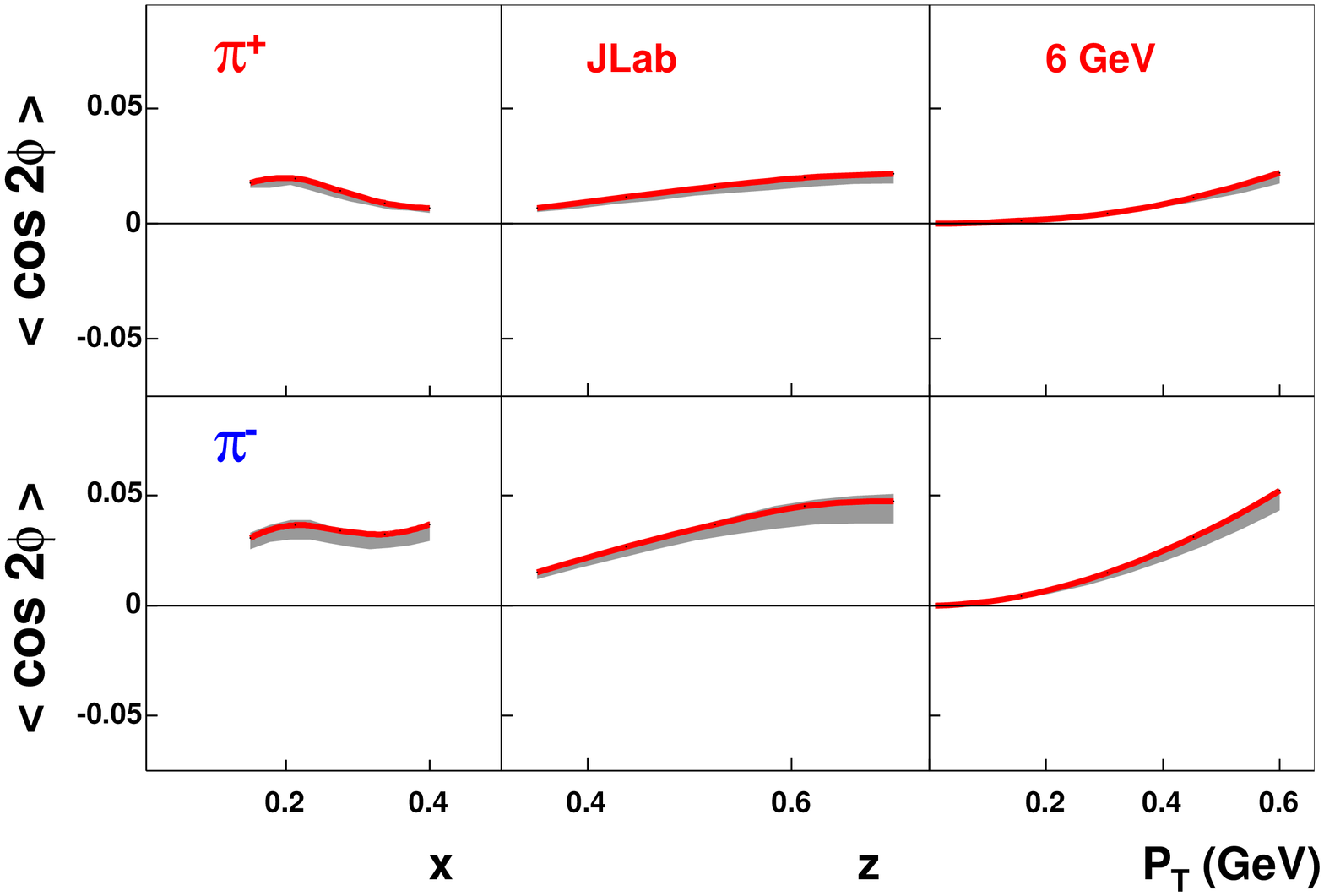}
\caption{\label{fig:jlab6_paper_udtest.ps}
Our prediction for $\langle \cos 2\phi \rangle$ asymmetry at JLab,  
with an incident beam energy of 6 GeV and a proton target 
using different $k_T$ widths for 
$u$ and $d$ quarks.
The bands correspond to a $\pm 50 \%$ variation 
of $\langle k_T^2 \rangle_d$ 
with respect to $\langle k_T^2 \rangle_u$, 
for which we take the value 0.25 GeV$^2$ 
(the upper end corresponds to $\langle k_T^2 \rangle_d = 0.375$ GeV$^2$, 
the lower end to $\langle k_T^2 \rangle_d = 0.125$ GeV$^2$).}
\end{figure}

Our prediction of a larger $\pi^-$ asymmetry as a signature 
of the Boer--Mulders effect is based on the assumption of 
a flavor--independent $k_T$-distribution. 
In order to check the robustness of this result, we 
varied the width of the Gaussian distribution 
for $d$ quarks (the $u$ distribution is well 
constrained by SIDIS data \cite{Anselmino:2005nn}), 
allowing it to be 50 \% larger or smaller than
the $u$ width, which is fixed to the value we used
above,  $\langle k_T^2 \rangle =0.25$ GeV$^2$.  
As one can see in Fig.~\ref{fig:jlab6_paper_udtest.ps}, 
even such a large difference between the $u$ and $d$ widths 
does not modify much the results.


\section{Conclusions}

The Boer--Mulders function $h_1^{\perp}$, 
one of the distributions describing 
the transverse spin and transverse momentum 
structure of the nucleon, is so far 
completely unknown. Its main effect
is an azimuthal $\cos 2\phi$ asymmetry 
in unpolarized SIDIS, an observable under an 
intense scrutiny by many ongoing and planned 
experiments. In this paper we presented some predictions 
for this asymmetry, taking all perturbative 
and non perturbative contributions into 
account. 
We found that $\langle \cos 2 \phi \rangle$ 
is generally of order of few percent, and in most 
cases is dominated, in the moderate $Q^2$ region,  
by a kinematical higher--twist effect arising 
from the intrinsic transverse motion of quarks, 
while the perturbative component is negligible.  
Concerning the Boer--Mulders mechanism, 
we showed that it is possible to learn about it 
by comparing
$\pi^+$ and $\pi^-$ production data, since
we predict that the contribution related to $h_1^{\perp}$
should be positive for $\pi^-$ and negative for 
$\pi^+$, which results in a $\pi^-$ asymmetry 
larger than the $\pi^+$ asymmetry. 

What emerges from our analysis is also 
the complementarity 
of the various experiments (HERMES, COMPASS, JLab). 
Taken altogether, the planned measurements 
of $\langle \cos 2 \phi \rangle$, with their 
variety of kinematical regimes and targets,  
will represent a very important piece of 
information on  
transverse spin and transverse momentum effects in 
the nucleon. 
  
\section{Acknowledgement}
This work is supported in part by the Italian Ministry 
of University and Research 
(PRIN2006) and by National Natural Science
Foundation of China (Nos.~10421503, 10575003, 10528510).

%

\end{document}